\documentclass[aps,floatfix,superscriptaddress,preprintnumbers,showpacs,showkeys]{revtex4}
\usepackage{amssymb}
\usepackage{amsmath}
\usepackage{amsfonts}
\usepackage{epsfig}
\usepackage{graphicx}
\usepackage{tabularx}
\usepackage{verbatim}
\newcommand{\eq}{\begin{eqnarray}}
\newcommand{\en}{\end{eqnarray}}

\begin{document}

\title{The interference effects of multi-channel pion-pion scattering contributions to the final states of $\Psi$- and $\Upsilon$-meson family decays}

\author{Yury S. Surovtsev}
\affiliation{Bogoliubov Laboratory of Theoretical Physics,
Joint Institute for Nuclear Research, 141980 Dubna, Russia}
\author{P. Byd\v{z}ovsk\'y}
\affiliation{Nuclear Physics Institute of the AS CR, 25068 \v{R}e\v{z},
Czech Republic}
\author{Thomas Gutsche}
\affiliation{Institut f\"ur Theoretische Physik,
Universit\"at T\"ubingen,
Kepler Center for Astro and Particle Physics,
Auf der Morgenstelle 14, D-72076 T\"ubingen, Germany}
\author{Robert Kami\'nski}
\affiliation{Institute of Nuclear Physics of the PAN, Cracow 31342, Poland}
\author{Valery E. Lyubovitskij}
\affiliation{Institut f\"ur Theoretische Physik,
Universit\"at T\"ubingen,
Kepler Center for Astro and Particle Physics,
Auf der Morgenstelle 14, D-72076 T\"ubingen, Germany}
\affiliation{Department of Physics, Tomsk State University,
634050 Tomsk, Russia}
\affiliation{Mathematical Physics Department,
Tomsk Polytechnic University,
Lenin Avenue 30, 634050 Tomsk, Russia}
\author{Miroslav Nagy}
\affiliation{Institute of Physics, SAS, Bratislava 84511, Slovak Republic}

\date{\today}

\begin{abstract}
 It is shown that the basic shape of dipion mass distributions in the two-pion transitions of both charmonia and bottomonia states are explained by an unified mechanism based on the contribution of the $\pi\pi$, $K\overline{K}$ and $\eta\eta$ coupled channels including their interference.
\end{abstract}



\maketitle

\section{Introduction}

In the analysis of practically all available data on two-pion transitions of the $\Upsilon$ mesons from the ARGUS, CLEO, CUSB, Crystal Ball, Belle, and {\it BaBar} Collaborations --- $\Upsilon(mS)\to\Upsilon(nS)\pi\pi$ ($m>n$, $m=2,3,4,5,$ $n=1,2,3$) --- the contribution of multi-channel $\pi\pi$ scattering in the final-state interactions is considered.
The analysis, which is aimed at studying the scalar mesons, is performed jointly considering the above bottomonia decays, the isoscalar S-wave processes $\pi\pi\to\pi\pi,K\overline{K},\eta\eta$, which are described in our model-independent approach based on analyticity and unitarity and using an uniformization procedure, and the charmonium decay processes ---
$J/\psi\to\phi\pi\pi$, $\psi(2S)\to J/\psi\pi\pi$ --- with data from the Crystal Ball, DM2, Mark~II, Mark~III, and BES~II Collaborations.

We show that the experimentally observed interesting (even mysterious) behavior of the $\pi\pi$ spectra of the $\Upsilon$-family decays, beginning from the second radial excitation and higher,  --- a bell-shaped form in the
near-$\pi\pi$-threshold region, smooth dips about 0.6~GeV in the $\Upsilon(4S,5S)\to\Upsilon(1S) \pi^+ \pi^-$, about 0.45~GeV in the $\Upsilon(4S,5S)\to\Upsilon(2S) \pi^+ \pi^-$, and about 0.7~GeV in the
$\Upsilon(3S)\to\Upsilon(1S)(\pi^+\pi^-,\pi^0\pi^0)$, and also sharp dips about 1~GeV
in the $\Upsilon(4S,5S)\to\Upsilon(1S) \pi^+ \pi^-$ --- is explained by the interference between the $\pi\pi$ scattering, $K\overline{K}\to\pi\pi$ and $\eta\eta\to\pi\pi$ contributions to the final states of these decays (by the constructive one in the near-$\pi\pi$-threshold region and by the destructive one in the dip regions).

\section{The effect of multi-channel $\pi\pi$ scattering in decays of the $\psi$- and $\Upsilon$-meson families}

Considering multi-channel $\pi\pi$ scattering, we shall deal with the 3-channel case  ($\pi\pi\to\pi\pi,K\overline{K},\eta\eta$) because it was shown \cite{SBLKN-jpgnpp14} that this is a minimal number of coupled channels needed for obtaining correct values of $f_0$-resonance parameters. When performing our combined analysis data for the multi-channel $\pi\pi$ scattering were taken from many papers (see Refs. in our paper~\cite{SBLKN-prd14}).
For the decay $J/\psi\to\phi\pi^+\pi^-$ data were taken from Mark III, DM2 and BES II Collaborations;
for $\psi(2S)\to J/\psi(\pi^+\pi^-~{\rm and}~\pi^0\pi^0)$ --- from Mark~II and Crystal Ball(80) (see Refs. also in~\cite{SBLKN-prd14}).
For $\Upsilon(2S)\to\Upsilon(1S)(\pi^+\pi^-~{\rm and}~\pi^0\pi^0)$ data were used from ARGUS~\cite{Argus}, CLEO~\cite{CLEO}, CUSB~\cite{CUSB}, and Crystal Ball~\cite{Crystal_Ball(85)} Collaborations; for
$\Upsilon(3S)\to\Upsilon(1S)(\pi^+\pi^-,\pi^0\pi^0)$ and $\Upsilon(3S)\to\Upsilon(2S)(\pi^+\pi^-,\pi^0\pi^0)$ --- from CLEO \cite{CLEO(94),CLEO07}; for $\Upsilon(4S)\to\Upsilon(1S,2S)\pi^+\pi^-$ --- from {\it BaBar} \cite{BaBar06} and Belle \cite{Belle}; for $\Upsilon(5S)\to\Upsilon(1S,2S,3S)\pi^+\pi^-$ --- from Belle Collaboration~\cite{Belle}.

The used formalism for calculating the di-meson mass distributions in the quarkonia decays is analogous to the one proposed in Ref.~\cite{MP-prd93} for the decays $J/\psi\to\phi(\pi\pi, K\overline{K})$ and $V^{\prime}\to V\pi\pi$ ($V=\psi,\Upsilon$) but with allowing for also amplitudes of transitions between the $\pi\pi$, $K\overline{K}$ and $\eta\eta$ channels in decay formulas. There was assumed that the pion pairs in the final state have zero isospin and spin. Only these pairs of pions undergo final state interactions whereas the final $\Upsilon(nS)$ meson ($n<m$) remains a spectator. The amplitudes of decays are related with the scattering amplitudes
$T_{ij}$ $(i,j=1-\pi\pi,2-K\overline{K},3-\eta\eta)$ as follows
\begin{eqnarray}
&&F\bigl(J/\psi\to\phi\pi\pi\bigr)=c_1(s)T_{11}+\Bigl(\frac{\alpha_2}{s-\beta_2}+c_2(s)\Bigr)T_{21}+c_3(s)T_{31},\\
&&F\bigl(\psi(2S)\to\psi(1S)\pi\pi\bigr)=d_1(s)T_{11}+d_2(s)T_{21}+d_3(s)T_{31},\\
&&F\bigl(\Upsilon(mS)\to\Upsilon(nS)\pi\pi\bigr) = e_1^{(mn)}T_{11}
+ e_2^{(mn)}T_{21}+ e_3^{(mn)}T_{31},\\
&&~~~~~~~~~~~~m>n,~ m=2,3,4,5,~ n=1,2,3\nonumber
\end{eqnarray}
where $c_i=\gamma_{i0}+\gamma_{i1}s$, $d_i=\delta_{i0}+\delta_{i1}s$ and $e_i^{(mn)}=\rho_{i0}^{(mn)}+\rho_{i1}^{(mn)}s$; indices $m$ and $n$ correspond to $\Upsilon(mS)$ and $\Upsilon(nS)$, respectively. The free parameters $\alpha_2$, $\beta_2$, $\gamma_{i0}$, $\gamma_{i1}$, $\delta_{i0}$, $\delta_{i1}$, $\rho_{i0}^{(mn)}$ and $\rho_{i1}^{(mn)}$ depend on the couplings of $J/\psi$, $\psi(2S)$ and the $\Upsilon(mS)$ to the channels $\pi\pi$, $K\overline{K}$ and $\eta\eta$. The pole term in eq.(1) in front of $T_{21}$ is an approximation of possible $\phi K$ states, not forbidden by OZI rules.

The amplitudes $T_{ij}$ are expressed through the $S$-matrix elements
~$S_{ij}=\delta_{ij}+2i\sqrt{\rho_1\rho_2}T_{ij}$
where $\rho_i=\sqrt{1-s_i/s}$ and $s_i$ is the reaction threshold. The $S$-matrix elements are parameterized on the uniformization plane of the $\pi\pi$ scattering amplitude by poles and zeros which represent resonances. The uniformization plane is obtained by a conformal map of the 8-sheeted Riemann surface, on which the three-channel $S$ matrix is determined, onto the plane. In the uniformizing variable used we have neglected the $\pi\pi$-threshold branch point and allowed for the $K\overline{K}$- and $\eta\eta$-threshold branch points and left-hand branch point at $s=0$ related to the crossed channels. The background is introduced to the amplitudes in a natural way: on the threshold of each important channel there appears generally speaking a complex phase shift. It is important that we have obtained practically zero background of the $\pi\pi$ scattering in the scalar-isoscalar channel. It confirms well our representation of resonances.

The expression ~$N|F|^{2}\sqrt{(s-s_1)[m_\psi^{2}-(\sqrt{s}-m_\phi)^{2}][m_\psi^2-
(\sqrt{s}+m_\phi)^2]}$~ for decay $J/\psi\to\phi\pi\pi$
and the analogues relations for $\psi(2S)\to\psi(1S)\pi\pi$~ and $\Upsilon(mS)\to\Upsilon(nS)\pi\pi$ give the di-meson mass distributions. $N$ (normalization to experiment) is: for $J/\psi\to\phi\pi\pi$ ~0.5172 (Mark III), 0.1746 (DM 2) and 3.8 (BES II); for $\psi(2S)\to J/\psi\pi^+\pi^-$ 1.746 (Mark II); for $\psi(2S)\to J/\psi\pi^0\pi^0$ 1.6891 (Crystal Ball(80)); for $\Upsilon(2S)\to \Upsilon(1S)\pi^+\pi^-$ 4.1758 (ARGUS), ~2.0445 (CLEO(94)) and 1.0782 (CUSB); for $\Upsilon(2S)\to\Upsilon(1S)\pi^0\pi^0$  0.0761 (Crystal Ball(85)); for $\Upsilon(3S)\to\Upsilon(1S)(\pi^+\pi^-~{\rm and}~\pi^0\pi^0)$  19.8825 and ~4.622 (CLEO(07)); for
$\Upsilon(3S)\to\Upsilon(2S)(\pi^+\pi^-$ ${\rm and}~\pi^0\pi^0)$ ~1.6987 and ~1.1803 (CLEO(94)); for $\Upsilon(4S)\to\Upsilon(1S)\pi^+\pi^-$ ~4.6827 ({\it BaBar}(06)) and ~0.3636 (Belle(07)); for $\Upsilon(4S)\to\Upsilon(2S)\pi^+\pi^-$, ~37.9877 ({\it BaBar}(06)); for $\Upsilon(5S)\to\Upsilon(1S)\pi^+\pi^-$,  $\Upsilon(5S)\to\Upsilon(2S)\pi^+\pi^-$ and $\Upsilon(5S)\to\Upsilon(3S)\pi^+\pi^-$ respectively ~0.2047, 2.8376 and 6.9251 (Belle(12)).

Satisfactory combined description of all considered processes (including $\pi\pi\to\pi\pi,K\overline{K},\eta\eta$) is obtained with the total $\chi^2/\mbox{ndf}=736.457/(710 - 118)\approx1.24$; for the $\pi\pi$ scattering, $\chi^2/\mbox{ndf}\approx1.15$.

Studying the decays of charmonia and bottomonia, we investigated the role of
the individual $f_0$ resonances in contributing to the shape of the dipion mass
distributions.
In this case we switched off only those resonances [$f_0(500)$, $f_0(1370)$,
$f_0(1500)$ and $f_0(1710)$], removal of which can be somehow compensated by
correcting the background to have
the more-or-less acceptable description of the multichannel $\pi\pi$ scattering.
First, when leaving out before-mentioned resonances, a minimal set of the $f_0$
mesons consisting of the $f_0(500)$, $f_0(980)$, and $f_0^\prime(1500)$ is
sufficient to achieve a description of the processes
$\pi\pi\!\to\!\pi\pi,K\overline{K},\eta\eta$ with a total $\chi^2/\mbox{ndf}\approx1.20$.
Second, from these three mesons only the $f_0(500)$ can be switched off while still
obtaining a reasonable description of multi-channel $\pi\pi$-scattering (though with
an appearance of the pseudo-background) with a total $\chi^2/\mbox{ndf}\approx1.43$.
In figures 1-3 the solid lines correspond to contribution of all relevant $f_0$-resonances; the dotted, of the $f_0(500)$, $f_0(980)$, and $f_0^\prime(1500)$; the dashed, of the $f_0(980)$ and $f_0^\prime(1500)$.
\begin{figure}[!h]
\begin{center}
\includegraphics[width=0.5\textwidth,angle=0]{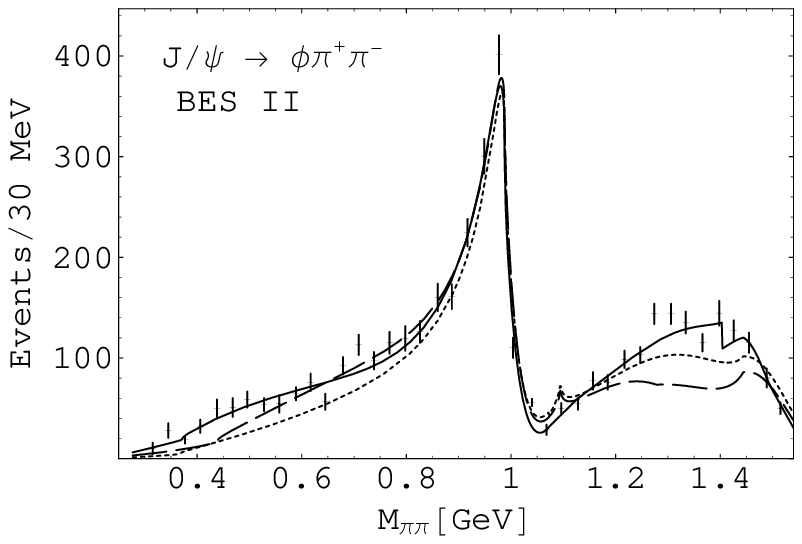}\\
\vspace*{0.3cm}
\includegraphics[width=0.45\textwidth]{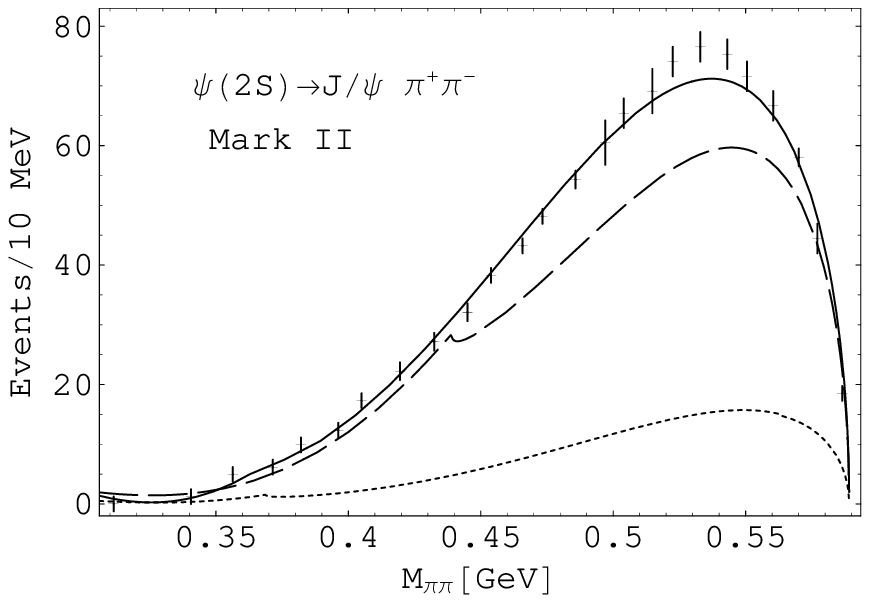}
\includegraphics[width=0.45\textwidth]{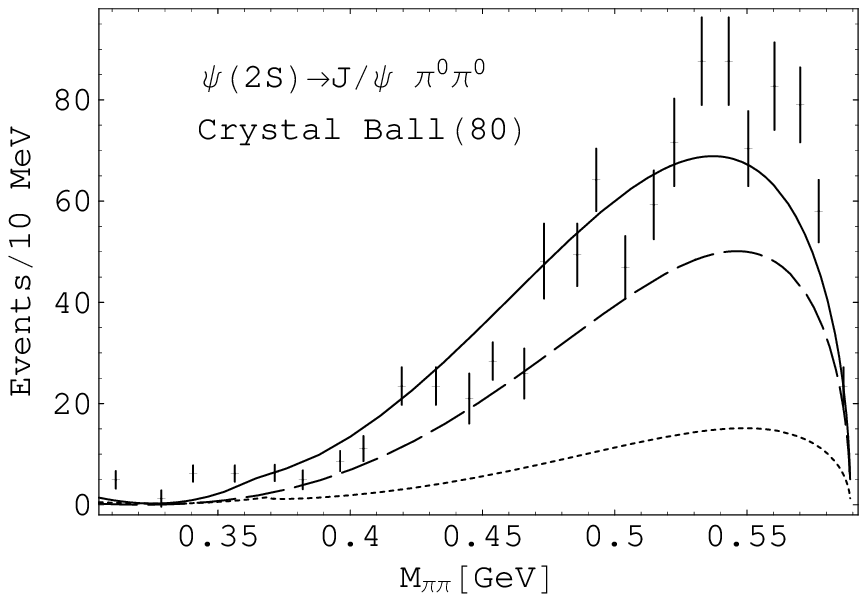}
\caption{The decays $J/\psi\to\phi\pi\pi$ and $\psi(2S)\to J/\psi\pi\pi$. The solid lines correspond to contribution of all relevant $f_0$-resonances; the dotted, of the $f_0(500)$, $f_0(980)$, and $f_0^\prime(1500)$; the dashed, of the $f_0(980)$ and $f_0^\prime(1500)$.}
\end{center}\label{fig:BESII}
\end{figure}
\begin{figure}[!htb]
\begin{center}
\includegraphics[width=0.45\textwidth]{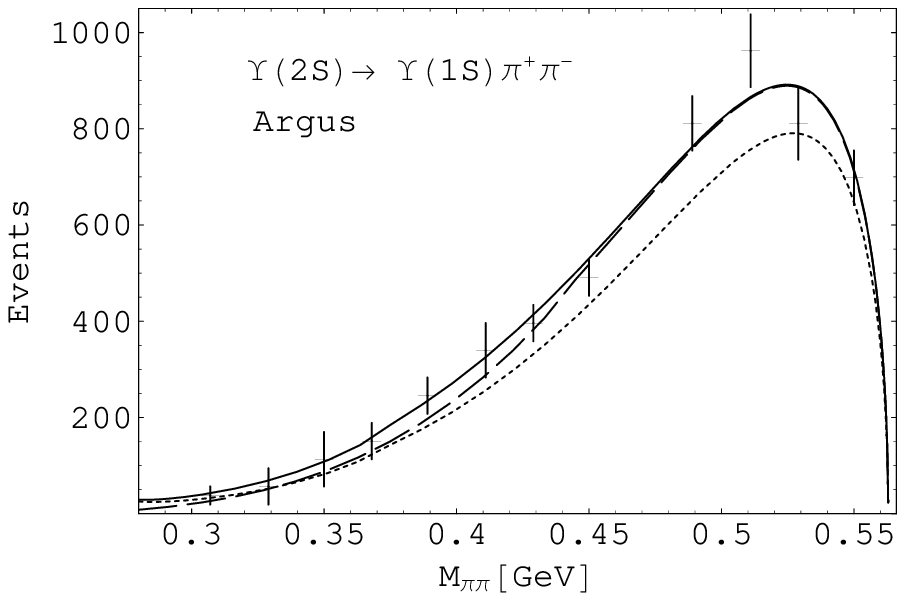}
\includegraphics[width=0.45\textwidth]{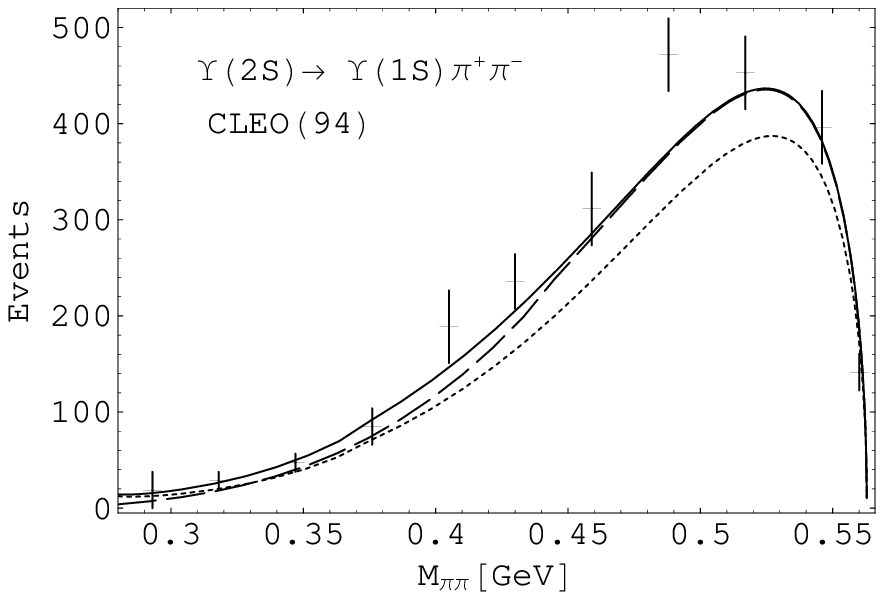}\\
\includegraphics[width=0.45\textwidth]{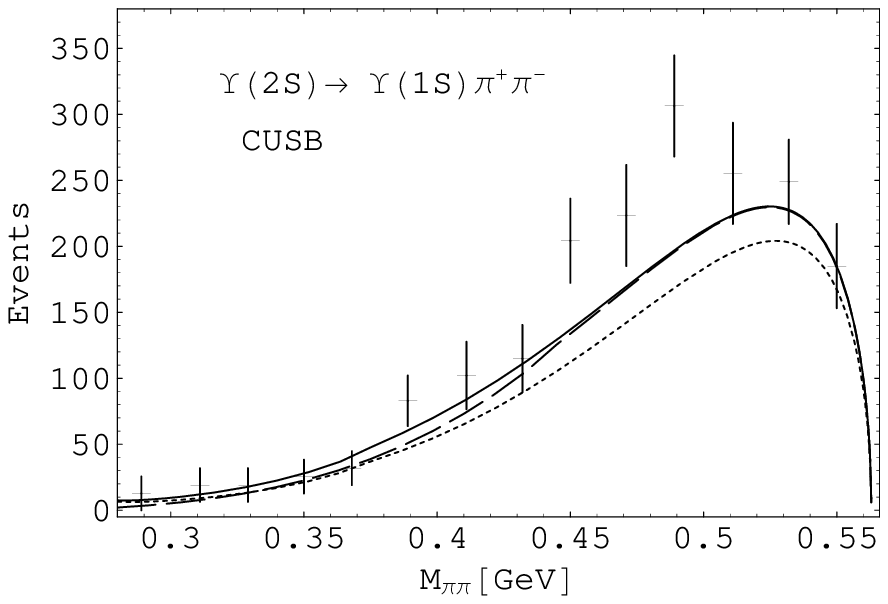}
\includegraphics[width=0.45\textwidth]{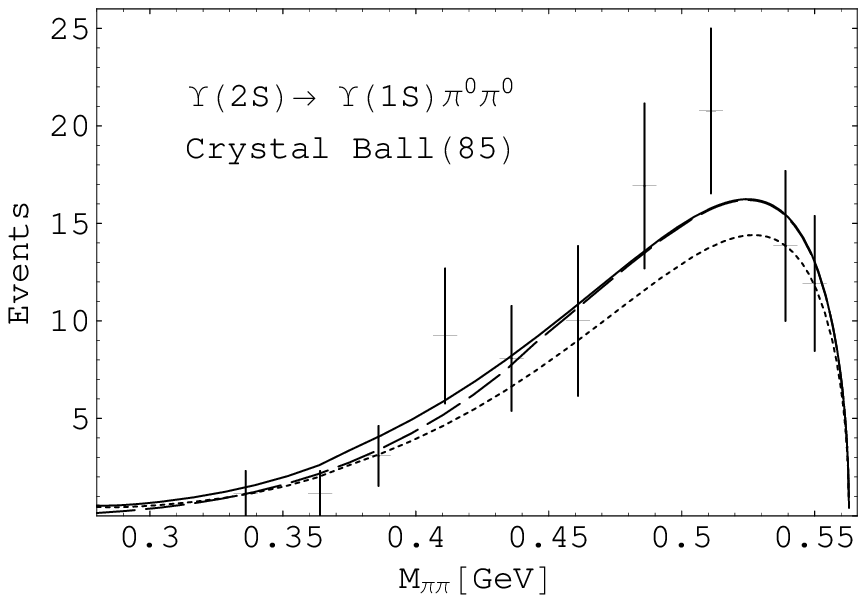}\\
\includegraphics[width=0.45\textwidth]{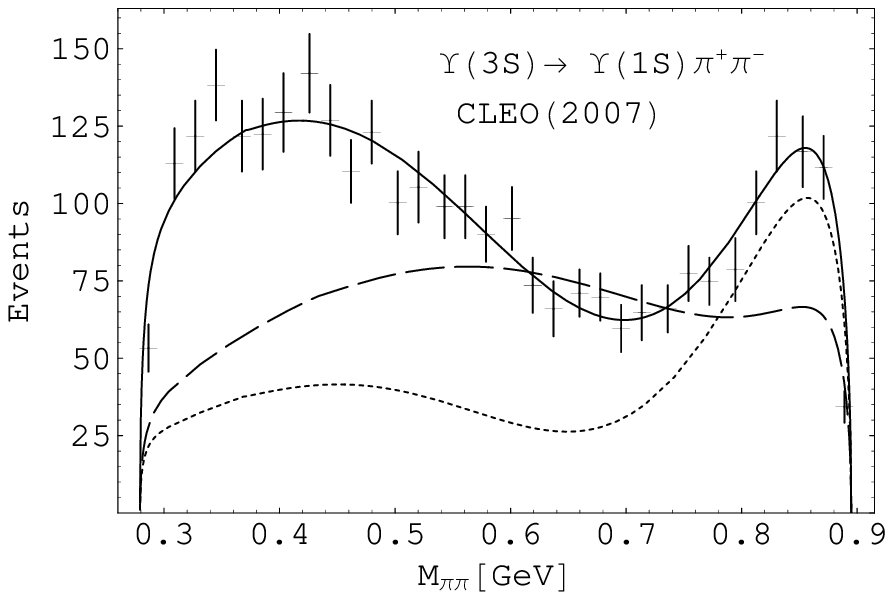}
\includegraphics[width=0.45\textwidth]{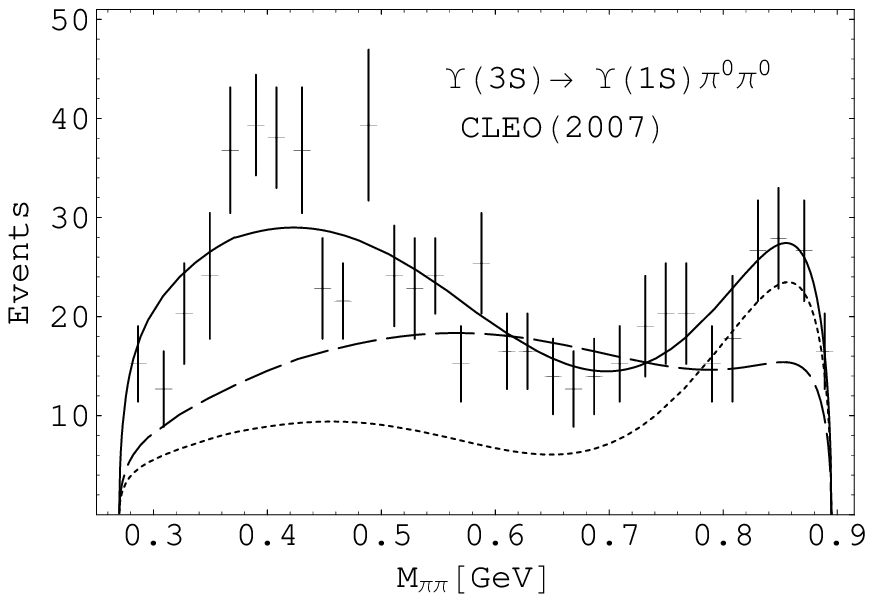}\\
\includegraphics[width=0.45\textwidth]{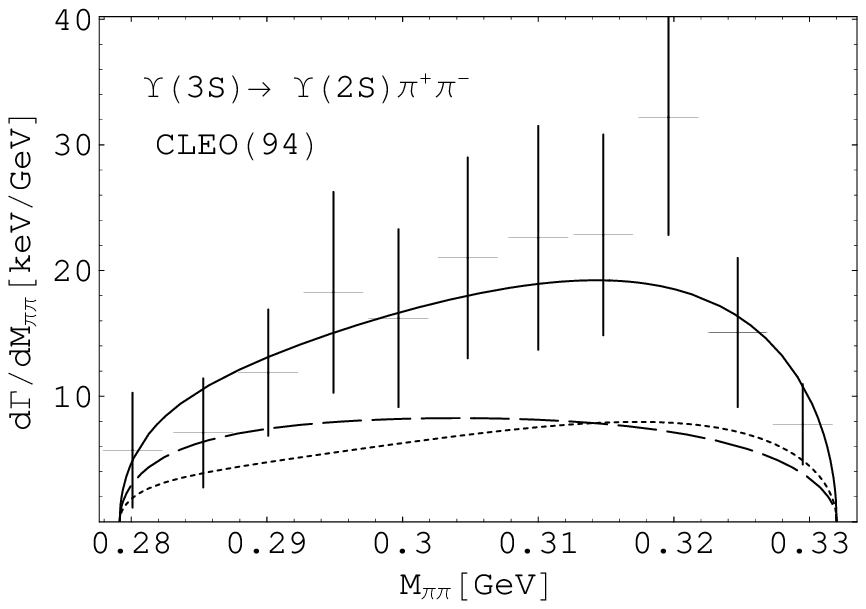}
\includegraphics[width=0.45\textwidth]{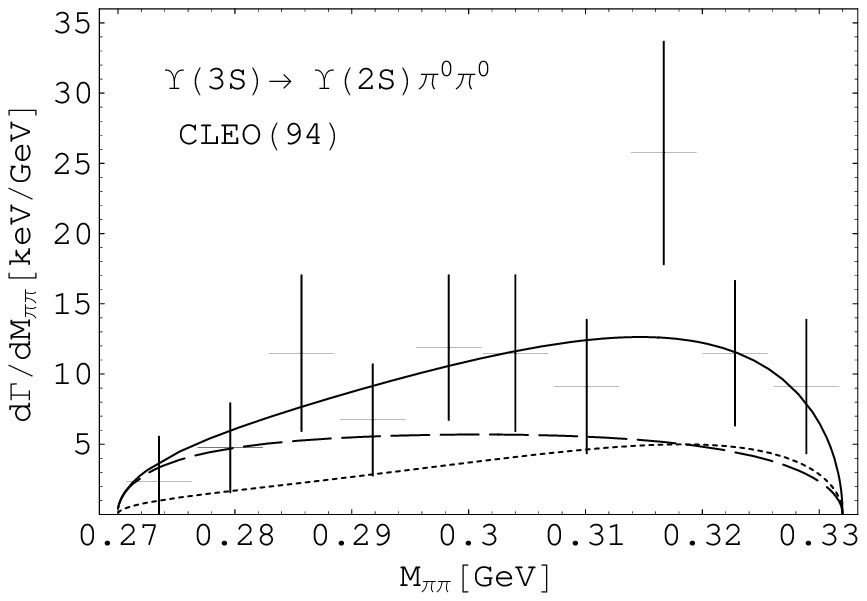}
\vspace*{-0.0cm}\caption{The decays $\Upsilon(2S)\to\Upsilon(1S)\pi\pi$ (two upper panels), $\Upsilon(3S)\to\Upsilon(1S)\pi\pi$ (middle panel) and $\Upsilon(3S)\to\Upsilon(2S)\pi\pi$ (lower panel). The solid lines correspond to contribution of all relevant $f_0$-resonances; the dotted, of the $f_0(500)$, $f_0(980)$, and $f_0^\prime(1500)$; the dashed, of the $f_0(980)$ and $f_0^\prime(1500)$.
}
\end{center}\label{fig:Ups21}
\end{figure}
\begin{figure}[!htb]
\begin{center}
\includegraphics[width=0.45\textwidth]{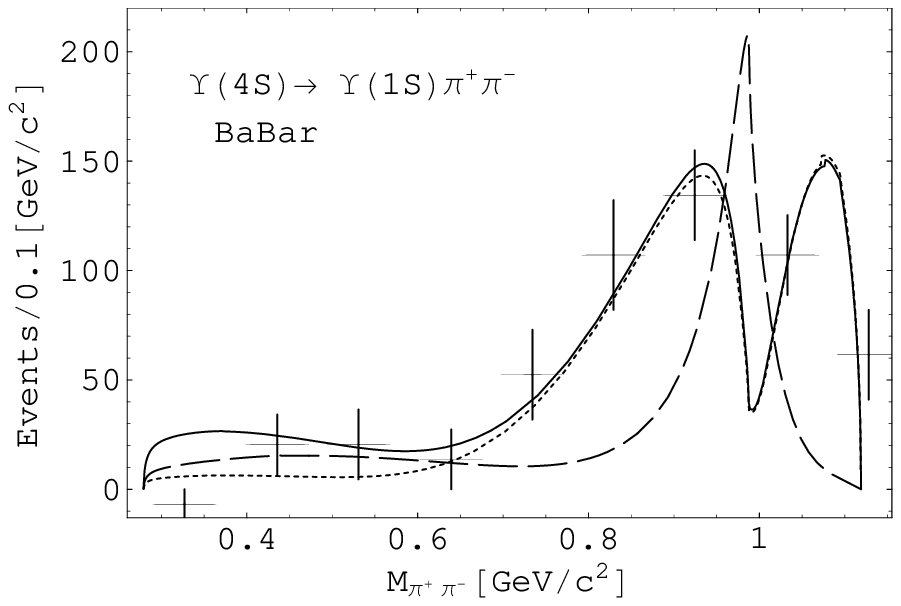}
\includegraphics[width=0.45\textwidth]{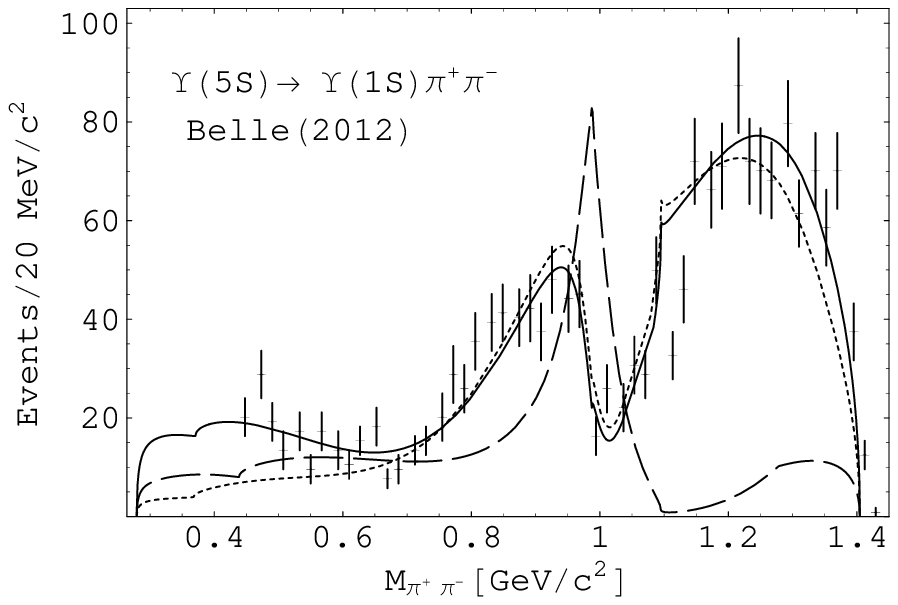}\\
\includegraphics[width=0.45\textwidth]{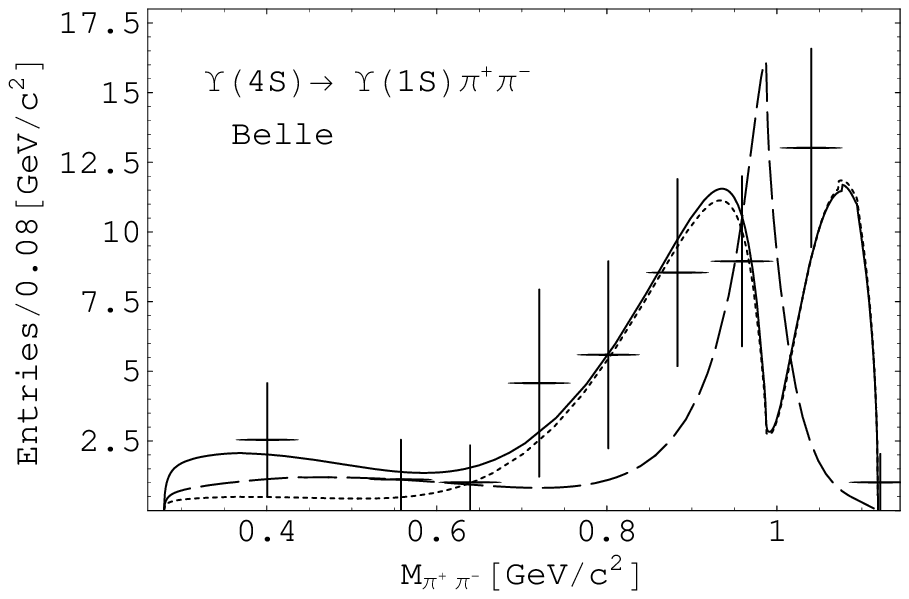}
\includegraphics[width=0.45\textwidth]{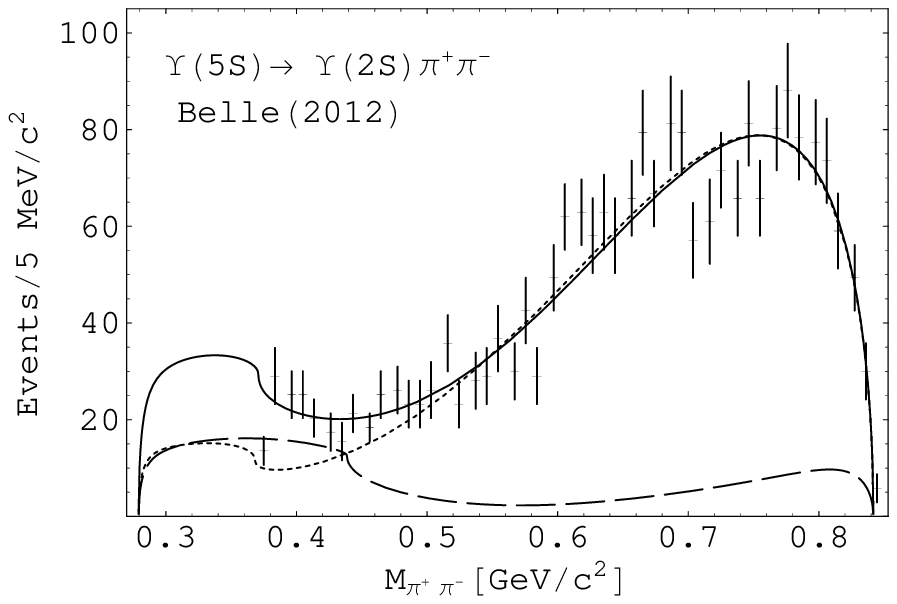}\\
\includegraphics[width=0.45\textwidth]{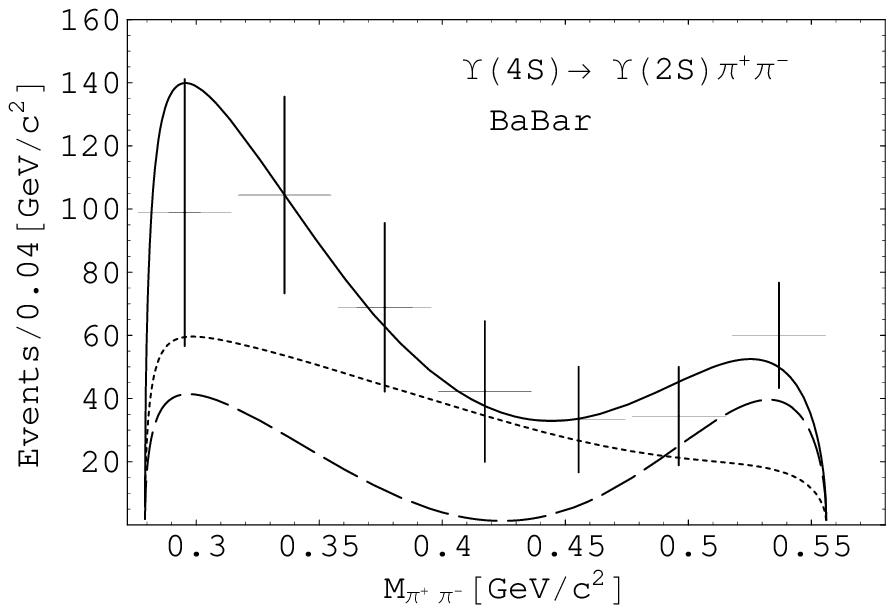}
\includegraphics[width=0.45\textwidth]{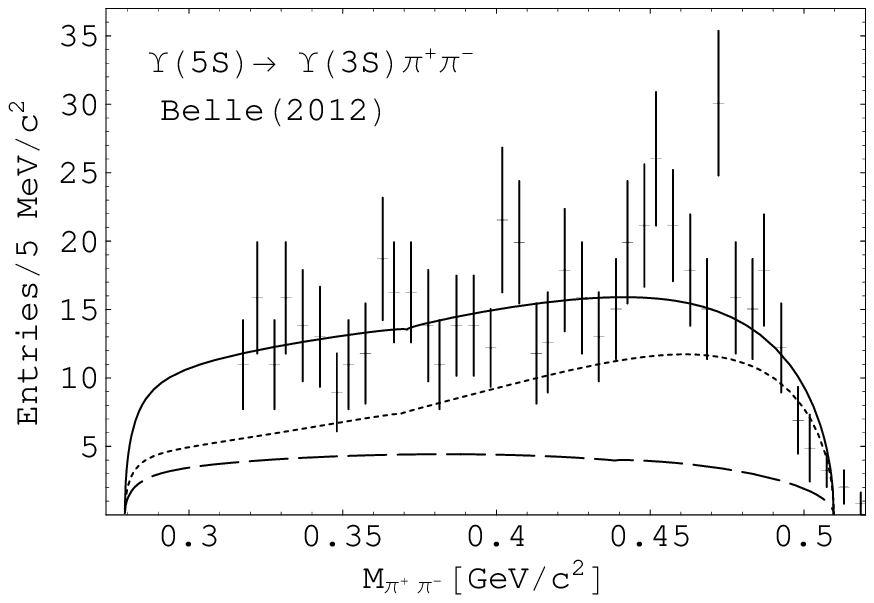}
\vspace*{-0.26cm}\caption{The decays $\Upsilon(4S)\to\Upsilon(1S,2S)\pi^+\pi^-$ (left-hand) and $\Upsilon(5S)\to\Upsilon(ns)\pi^+ \pi^-$ ($n=1,2,3$) (right-hand). The solid lines correspond to contribution of all relevant $f_0$-resonances;
the dotted, of the $f_0(500)$, $f_0(980)$, and $f_0^\prime(1500)$;
the dashed, of the $f_0(980)$ and $f_0^\prime(1500)$.}
\end{center}
\end{figure}

\section{Conclusions}

The combined analysis was performed for data on isoscalar S-wave processes
$\pi\pi\to\pi\pi,K\overline{K},\eta\eta$ and on the decays of the charmonia ---
$J/\psi\to\phi\pi\pi$, $\psi(2S)\to J/\psi\,\pi\pi$ --- and of the bottomonia ---
$\Upsilon(mS)\to\Upsilon(nS)\pi\pi$ ($m>n$, $m=2,3,4,5,$ $n=1,2,3$) from the ARGUS,
Crystal Ball, CLEO, CUSB, DM2, Mark~II, Mark~III, BES~II, {\it BaBar}, and Belle
Collaborations.

It is shown that the dipion mass spectra in the above-indicated decays of charmonia
and bottomonia are explained by the unified mechanism which is based on our previous
conclusions on wide resonances \cite{SBLKN-jpgnpp14,SBLKN-prd14} and is related
to contributions of the $\pi\pi$, $K\overline{K}$ and $\eta\eta$ coupled channels
including their interference. It is shown that in the final states of these decays (except $\pi\pi$ scattering) the contribution of coupled processes, e.g., $K\overline{K},\eta\eta\to\pi\pi$, is important even if these processes are energetically forbidden.

The role of the individual $f_0$ resonances in making up the shape of the dipion mass
distributions in the charmonia and bottomonia decays is considered. Note the unexpected result --- a considerable contribution of the $f_0(1370)$ to the bell-shaped form of
the dipion mass spectra of bottomonia decays in the near-$\pi\pi$-threshold region.

Since describing the bottomonia decays, we did not change resonance parameters in
comparison with the ones obtained in the combined analysis of the processes
$\pi\pi\to\pi\pi,K\overline{K},\eta\eta$ and charmonia decays, the results of this
analysis confirm all of our earlier conclusions on the scalar mesons \cite{SBLKN-jpgnpp14}.\\

This work was supported in part by the Heisenberg-Landau Program, by the Votruba-Blokhintsev Program for Cooperation of Czech Republic with JINR, by the Grant Agency of the Czech Republic (grant No. P203/15/04301), by the Grant Program of Plenipotentiary of Slovak Republic at JINR, by the Bogoliubov-Infeld Program for Cooperation of Poland with JINR, by the BMBF (Project 05P2015, BMBF-FSP 202), by Tomsk State University Competitiveness Improvement Program, the Russian Federation program ``Nauka'' (Contract No.\ 0.1526.2015, 3854), by Slovak Grant Agency VEGA under contract No.2/0197/14, and by the Polish National Science Center (NCN) grant DEC-2013/09/B/ST2/04382.



\begin{thebibliography}{999}

\bibitem{SBLKN-jpgnpp14}
Yu.S.~Surovtsev {\it et al.}, J.\ Phys.\ G\: Nucl.\ Part.\ Phys.\ \textbf{41}, 025006 (2014);
Phys.\ Rev.\ D \textbf{86}, 116002 (2012).

\bibitem{SBLKN-prd14}
Yu.S.~Surovtsev {\it et al.}, Phys.\ Rev.\ D \textbf{89}, 036010 (2014).

\bibitem{Argus}
H.~Albrecht {\it et al.} (ARGUS Collaboration), Phys.\ Lett.\ \textbf{134B}, 137 (1984).

\bibitem{CLEO}
D.~Besson {\it et al.} (CLEO Collaboration), Phys.\ Rev.\ D \textbf{30}, 1433 (1984).

\bibitem{CUSB}
V.~Fonseca {\it et al.} (CUSB Collaboration), Nucl.\ Phys.\ B \textbf{242}, 31 (1984).

\bibitem{Crystal_Ball(85)}
D.~Gelphman {\it et al.} (Crystal Ball Collaboration), Phys.\ Rev.\ D \textbf{32}, 2893 (1985).

\bibitem{CLEO07}
D.~Cronin-Hennessy {\it et al.} (CLEO Collaboration), Phys.\ Rev.\ D \textbf{76}, 072001 (2007).

\bibitem{CLEO(94)} F.~Butler {\it et al.} (CLEO Collaboration), Phys.\ Rev.\ D \textbf{49}, 40 (1994).

\bibitem{BaBar06}
B.~Aubert {\it et al.} (BaBar Collaboration), Phys.\ Rev.\ Lett. \textbf{96}, 232001 (2006).

\bibitem{Belle}
A.~Sokolov {\it et al.} (Belle Collaboration), Phys.\ Rev.\ D \textbf{75}, 071103 (2007);
A.~Bondar {\it et al.} (Belle Collaboration), Phys.\ Rev. \ Lett. \textbf{108}, 122001 (2012).

\bibitem{MP-prd93}
D.~Morgan and M.R.~Pennington, Phys.\ Rev.\ D \textbf{48}, 1185, 5422 (1993).

\end{thebibliography}
\end{document}